\documentclass[aps,prb,twocolumn,groupedaddress]{revtex4}

\usepackage{amsmath}
\usepackage{graphics}
\usepackage{graphicx}
\usepackage{amsmath}
\usepackage{amsfonts}
\usepackage{amssymb}

\def \beq{\begin{equation}}
\def \eeq{\end{equation}}
\def \beqar{\begin{eqnarray}}
\def \eeqar{\end{eqnarray}}

\begin{document}

\title{\bf{Lower critical field and intragrain critical current density
in the ruthenate-cuprate
RuSr$_{2}$Gd$_{1.5}$Ce$_{0.5}$Cu$_{2}$O$_{10}$}}

\author{M.G. das Virgens$^{1,2}$, S. Garc\'{\i}a$^{1,3}$, L. Ghivelder$^1$}

\affiliation{$^1$Instituto de F\'{\i}sica, Universidade Federal do
Rio de Janeiro, C.P. 68528, Rio de Janeiro, RJ 21941-972,
Brazil\\$^2$Instituto de F\'{\i}sica, Universidade Federal Fluminense, C.P. 68528, Niter%
\'{o}i, RJ 21945-970, Brazil\\$^3$Laboratorio de
Superconductividad, Facultad de F\'{\i}sica-IMRE, Universidad de
La Habana, San L\'{a}zaro y L, Ciudad de La Habana 10400, Cuba}



\begin{abstract}
The lower critical field of the grains, $H_{c1}$, and the intragrain
critical current density, $J_{c}$, were determined for the superconducting
ruthenate-cuprate RuSr$_{2}$Gd$_{1.5}$Ce$_{0.5}$Cu$_{2}$O$_{10-\delta }$
[Ru-1222(Gd)] through a systematic study of the hysteresis in
magnetoresistance loops. A reliable method, based on the effects of the
magnetization of the grains on the net local field at the intergranular
junctions is provided, circumventing the problem of the strong masking of
the superconducting diamagnetic signal by the ferromagnetic background. The
temperature dependency of $H_{c1}$ and $J_{c}$ both exhibit a smooth
increase on cooling without saturation down to $T/T_{SC}$ $\cong $ 0.2. The
obtained $H_{c1}$ values vary between 150 and 1500 Oe in the 0.2 $\leq $ $%
T/T_{SC}$ $\leq $ 0.4 interval, for samples annealed in an oxygen flow;
oxygenation under high pressure (50 atm) leads to a further increase. These
values are much larger than the previously reported rough assessments (25-50
Oe), using conventional magnetization measurements. High $J_{c}$ values of $%
\sim $ 10$^{7}$ A/cm$^{2}$, comparable to the high-T$_{c}$ cuprates, were
obtained. The $H_{c1}(T)$ and $J_{c}(T)$ dependencies are explained in the
context of a magnetic phase separation scenario.
\end{abstract}

\maketitle

\section{Introduction}

Since the discovery of the so called magnetic superconductors RuSr$_{2}$RCu$%
_{2}$O$_{8}$ (Ru-1212) and
RuSr$_{2}$(R,Ce)$_{2}$Cu$_{2}$O$_{10-\delta }$ (Ru-1222), with
R=Gd, Eu,\cite{Bauernfeind,Tallon01} considerable effort has been
devoted to the understanding of the interplay between the
ferromagnetic (FM) component, emerging from the long-range order
of the Ru moments, and the onset of the superconducting (SC)
state.\cite{cond-mat} Among several important topics, the
possibility of $\pi $-phase formation across the RuO$_{2}$
layers,\cite{Pickett,GarciaMoss,Nachtrab} the itinerant
or localized character of the magnetism of the Ru moments,\cite%
{McCrone01,Tallon02,McCrone02} the magnetic phase separation scenario of
nanoscale FM clusters with superconductivity nucleating only in the
surrounding antiferromagnetic matrix,\cite{Xue01,Xue02} and the possibility
of triplet paring,\cite{Tallon02} have been considered to explain how this
puzzling coexistence may occur. On the other hand, some important
superconducting parameters have been less thoroughly investigated. Reports
include the determination of the coherence length $\xi $ and the higher
critical field $H_{c2}$,\cite{Escote,Cimberle,Attanasio} and the intragrain
London penetration length $\lambda _{L}$,\cite{Xue02,Xue03,GarciaIV,Lorenz}
and rough estimations of the lower critical field of the grains, $H_{c1}$.%
\cite{Awana} In relation to the determination of $H_{c1}$, a diamagnetic
signal has been observed in a few cases at the low field range of the $M(H)$
magnetization loops in Ru-1222, with a negative minimum at about 25 Oe\cite%
{Awana} and 50 Oe.\cite{Felner01,Felner02}

There are also fewer studies on relevant intrinsic superconducting
properties, such as the intragrain critical current density, $J_{c}$,\cite%
{Felner IJMP} mainly because the strong FM contribution to the
magnetization from the Ru sublattice makes impracticable the use
of the magnetic hysteresis loops to determine both $H_{c1}$ and
$J_{c}$. In the present study we overcome this intrinsic
difficulty and present a reliable method to determine these
magnitudes and their temperature dependencies in Ru-1222(Gd),
through a systematic study of the hysteresis in the isotemperature
magnetoresistance $R(H)$ curves. Since no single crystals are
available for this compound, polycrystalline materials were used
in the present investigation. Two different Ru-1222(Gd) samples
were studied, obtained under different partial oxygen pressures.
At variance with the behavior of the high-T$_{c}$ cuprates, a
monotonic increase without saturation in both $H_{c1}$ and $J_{c}$
on cooling was observed, reaching
values as high as $H_{c1}$ $\sim $ 1000 Oe and $J_{c}$ $\sim $ 10$^{7}$ A/cm$%
^{2}$, at $T$ = 7.5 K. A comparison with YBa$_{2}$Cu$_{3}$O$_{7}$ (YBCO) and
with the results reported for modeling the magnetic properties of Ru-1212 on
the basis of the theory of the SC/FM multilayers is presented.

\section{Experimental}

Polycrystalline RuSr$_{2}$Gd$_{1.5}$Ce$_{0.5}$Cu$_{2}$O$_{10-\delta }$ was
prepared by conventional solid-state reaction using an oxygen flow in the
final heat treatment. The room temperature x-ray diffraction pattern
corresponds to Ru-1222(Gd), with no spurious lines. Scanning electron
microscopy revealed a dense grain packing, with an average grain size d $%
\cong $ 0.5-1 $\mu $m. More details on sample preparation and microstructure
can be found elsewhere.\cite{GarciaMR} After characterization, the
as-prepared (\textit{asp}) sample was annealed for 24 hours at 600$^{o}$C
under 50 atm of pure oxygen [high oxygen pressure (\textit{hop}) sample].
Magnetotransport and ac magnetic susceptibility measurements\cite{GarciaMR}
reproduce the behavior of good quality samples.\cite{Tallon02} Bars of $%
\cong $\ 10 mm in length and 0.6 mm$^{2}$ cross sectional area were cut from
the sintered pellet. The resistance was measured with a standard four
contacts probe using a Quantum Design PPMS system, at $T$ = 7.5, 8, 9, 10,
11.25, 12.5, and 15 K for the \textit{asp} sample; the \textit{hop} sample
was also measured at 8.5 K. A large number of $R(H)$ curves were collected
for each temperature to accurately follow the different characteristic
regimes found in the magnetoresistance response (as described below), and to
warrant a reliable quantitative determination of the fields at which the
transition from one regime to another occur. The most relevant experimental
parameter varied experimentally is the maximum applied field within the $%
R(H) $ curves, $H_{max}$, which ranged from a few tens of Oe up to 60 kOe.

\section{Results}

In order to characterize the transport behavior of the studied
samples we initially measured the temperature dependence of the
resistivity. It is clear from the data shown in Fig. 1 that
oxygenation under pressure strongly reduces the absolute
resistivity values, enlarges the linear behavior of the
normal-state region, and reduces the width of the SC transition.
In addition, the superconducting transition temperature,
$T_{SC}\cong 45K$, as determined from the peaks in the derivative
of the resistivity for the as prepared (\textit{asp}) sample,
increases by approximately 3 K in the \textit{hop} sample. These
features agree with previous reports in good quality
materials.\cite{McCrone01,Tallon02} The present
\begin{figure}[h!]
 \begin{center}
  \includegraphics[scale=0.6]{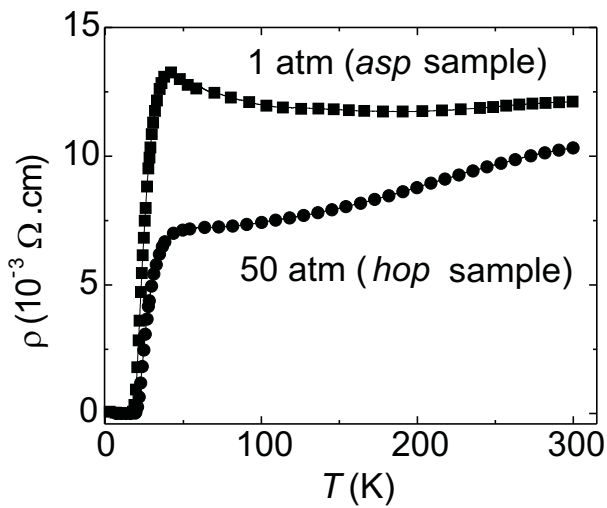}
  \caption{{\small{\it Temperature dependence of the resistivity for the Ru-1222 samples
obtained under oxygen pressure of 1atm (\textit{asp} sample) and
under high oxygen pressure of 50 atm (\textit{hop} sample).}}}
 \end{center}
\end{figure}\\
study concentrates in results of resistance as a function of
field, $R(H)$, measured at fixed temperature with different
maximum fields, $H_{max}$, in each sweep. Four different behaviors
were identified in the $R(H)$ curves as $H_{max}$ is increased: a)
a zero resistance region, typically for $H_{R=0}$ $<$ 100 Oe; b)
an interval of reversible dissipation up to an irreversible field
$H_{irr}$, from about 150 Oe to 1500 Oe for the \textit{asp}
sample and up to 3000 Oe for the \textit{hop} sample; c) an
$H_{max}$-dependent hysteretic behavior, and d) an hysteretic
response independent of $H_{max}$.
Figure 2 shows selected magnetoresistance hysteresis loops for the \textit{%
hop} sample with $H_{max}$ = 5000 Oe, measured at $T$ = 7.5 and 10 K,
normalized to the $R(H_{max})$ values. For each temperature, loops similar
in shape but broader were obtained for the \textit{asp} sample. In the
hysteretic region the resistance curves measured with decreasing external
field, $R(H\downarrow )$, are always below the corresponding virgin curves, $%
R(H\uparrow )$, measured with increasing field. When the difference $\Delta
R(H)=R(H\uparrow )-R(H\downarrow )$ is plotted, a peak is always obtained at
a certain intermediary value of the applied field, $H_{ext}$. With the rise
in $H_{max}$, the peak gradually increases its amplitude, $A_{P}(H_{max})$,
and shifts to higher $H_{ext}$ values until a certain field $H_{max}$ $%
\equiv $ $H_{sat}$ is reached, that we denote as a saturation field. For $%
H_{max}$ $>$ $H_{sat}$ the peak in $\Delta R(H)\ $remains
unchanged. To illustrate this behavior, Fig. 3 plots $\Delta R(H)\
$ as a function of $H_{ext}$\ for different values of $H_{max}$, for the \textit{hop%
} sample at $T$ = 12.5 K.

\begin{figure}[h!]
 \begin{center}
  \includegraphics[scale=0.53]{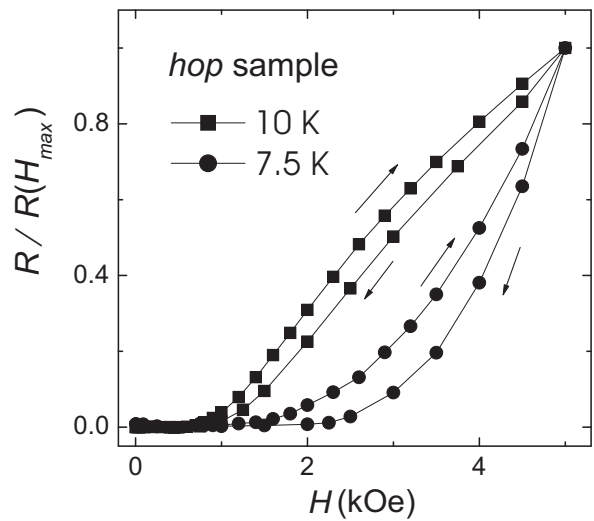}
  \caption{{\small{\it Selected isothermal hysteresis loops of the magnetoresistance of
Ru-1222 \textit{hop} sample, measured with $H_{max}$ = 5 kOe at
$T$ = 7.5 and 10 K. The arrows indicate the field direction during
the measurement. The curves are normalized to the maximum
resistance value, $R(H_{max})$.}}}
 \end{center}
\end{figure}

We focus our attention on a careful determination of $H_{irr}$ and
$H_{sat}$. Since $R(H\downarrow )$ diverge smoothly from
$R(H\uparrow )$, it is necessary to implement a method for this
evaluation. We proceeded as follows: a) the error $\delta R(H)$ in
the determination of $R(H)$ was taken as the average of the
standard deviations of each measured experimental point in a given
loop; we found that successive $R(H)$ curves measured at each
temperature for different $H_{max}$ (typically 15-20 curves)
always fell inside this $\delta R(H)$ criterion; b) for each
temperature, $A_{P}$ was plotted as a function of $H_{max}$, as
shown in Fig. 4 for the \textit{hop} sample
\begin{figure}[ht!]
 \begin{center}
  \includegraphics[scale=0.55]{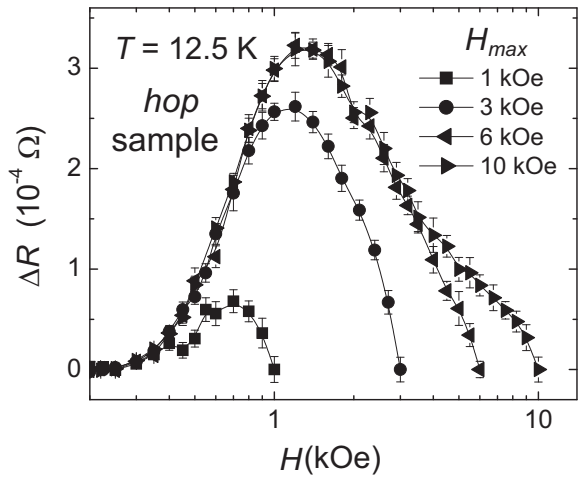}
  \caption{{\small{\it Field dependence of the difference between the virgin and the
decreasing field magnetoresistance curves, $\Delta
R(H)=R(H\uparrow )-R(H\downarrow )$, measured at $T$ = 12.5 K for
Ru-1222 \textit{hop} sample with different $H_{max}$. The peak
shifts to higher fields and increases its amplitude with the rise
in $H_{max}$ up to a certain saturation field $H_{sat}$.}}}
 \end{center}
\end{figure}\\at $T$ = 12.5 K, and the data fitted using a
polynomial function; the error in the determination of $A_{P}$ is
$2\delta R(H)$; c) $R(H\downarrow )$ is considered reversible if
$A_{P}$($H_{max}$) does not exceed the $\delta R(H)$ threshold, as
indicated by a dotted line above the zero resistance level in the
inset of Fig. 4 (low $H_{max}$ region); the irreversibility field,
$H_{irr}$, is determined from the intersection of the fitted
polynomial function with the $\delta R(H)$ level (Fig. 4, inset);
d) the plateau observed for $A_{P}(H_{max})$ for the high
$H_{max}$ range was fitted by a straight line parallel to the
$H_{max}$ axis (continuous line in Fig. 4, main panel); e) The
saturation field, $H_{sat}$, was determined as the field for which
the polynomial function diverges in $2\delta R(H)$ from the fitted
straight line, (intersection with the dashed line, parallel and
below the plateau level in Fig. 4, main panel). As we discuss below, $%
H_{irr} $\ is a good estimation of the lower critical field of the
grains, $H_{c1}$.

The temperature dependencies of $H_{c1}$ and $H_{sat}$ for the
\textit{asp} and \textit{hop} samples are plotted,in Fig. 5 and in
the inset of Fig. 6, respectively. The behavior of $H_{c1}$ for
YBCO (with the field parallel to the sample's \textit{c} axis) is
also shown for comparison in the inset of Fig. 5. The $H_{c1}(T)$
and $H_{sat}(T)$ curves for both samples monotonically decrease
with the increase in temperature and smoothly merge at about 15 K.
The values of $J_{c}$ at different temperatures, determined by
using the Bean critical state model,\cite{Bean} as described
below, are shown in Fig. 6. A systematic increase of $J_{c}$ on
cooling is found, with slightly higher values for the \textit{hop}
sample.

\section{Discussion}

A key point to understand the features of the $R(H)$ curves is to
recognize that the magnetization in the grains contributes to the
effective local field at the intergranular junctions. Hysteresis
in the magnetoresistance was recently observed in ceramic Ru-1222
samples,\cite{Felner03} and a preliminary analysis evidenced that
the granular character of the material is relevant. Unlike
Ru-1212, in Ru-1222 the coercive field and the remanent
magnetization vanish at the SC transition, $T_{SC}$ $\cong $ $45$
$K$, for both Gd\cite{Awana,Cardoso} and
Eu\cite{Felner02,Sonin,Xue04} based compounds, even though the
magnetic transition temperatures are much higher. Since the Gd
ions are decoupled from the Ru moments,\cite{Lee,Felner01,Knee} no
changes in the irreversibility behavior are observed when Gd is
replaced in Ru-1222 by Eu. The essential feature governing the
irreversibility response in Ru-1222 is the thick (R,Ce)$_{2}$O$_{2}$%
-fluorite-type blocks separating the RuO$_{2}$\ planes. The large separation
distance strongly weakens the superexchange interactions between the
magnetic layers, which are weakly coupled through dipole-dipole interaction.%
\cite{Zivkovic,GarciaMR}\ Therefore, the hysteretic behavior of the $R(H)$
curves has a SC origin in Ru-1222.
\begin{figure}[h!]
 \begin{center}
  \includegraphics[scale=0.6]{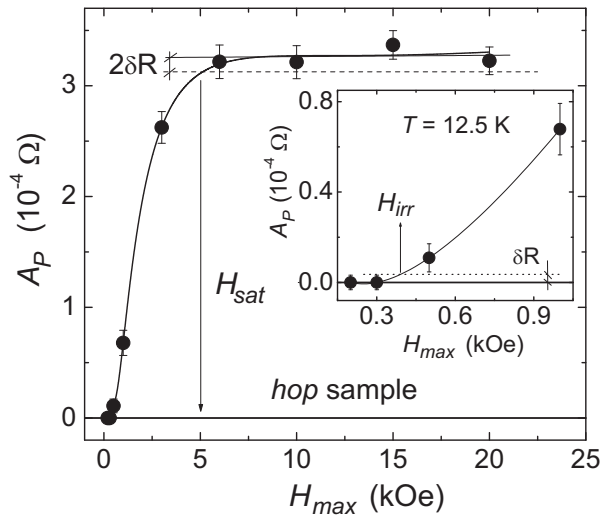}
  \caption{{\small{\it Amplitude $A_{P}$ of the peak in $\Delta R(H)=R(H\uparrow
)-R(H\downarrow )$, plotted as a function of the maximum applied
field $H_{max}$ for Ru-1222 \textit{hop} sample at $T$= 12.5 K.
The continuous line is a polynomial fitting. The straight line is
a fit of the experimental points in the plateau observed in the
high $H_{max}$ range. The dashed line shows the error in the
determination of $A_{P}$, measured from the plateau level. The
saturation field $H_{sat}$ is indicated. The inset shows an
enlarged section of the low $H_{max}$ region. The dotted line
marks the error level in the determination of the
magnetoresistance. The irreversible field $H_{irr}$ is
indicated.}}}
 \end{center}
\end{figure}

For a low applied magnetic field, $H_{ext}$, the net local field
at the intergranular junctions is not strong enough to destroy all
the SC percolative paths across the weak-linked network, leading
to the observed zero resistance region in $R(H)$. For $H_{ext}$
$>$ $H_{R=0}$, the connectivity of the network is affected in such
extent that the current across the sample can not flow without
dissipation, and a sizable resistance appears. The existence of a
reversible
\begin{figure}[h!]
 \begin{center}
  \includegraphics[scale=0.47]{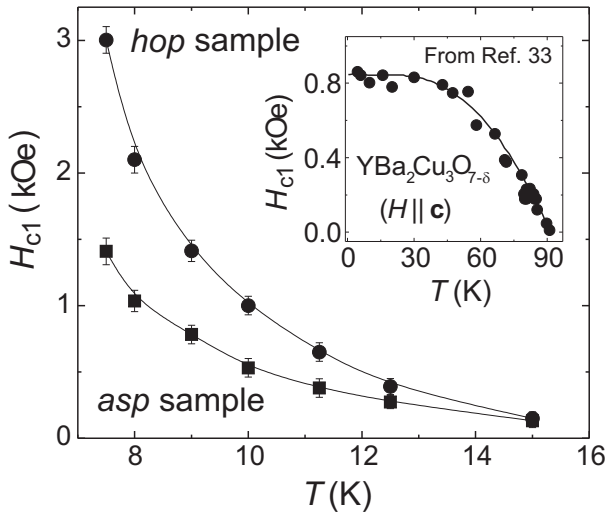}
  \caption{{\small{\it Temperature dependence of the lower critical field of the grains $%
H_{c1}$ for both Ru-1222 \textit{asp} and \textit{hop} samples.
The behavior of $H_{c1}$ (parallel to the \textit{c} axis) for
YBa$_{2}$Cu$_{3}$O$_{7}$ , obtained from Ref \cite{Wu} is also
shown for comparison. The continuous lines are guides to the
eyes.}}}
 \end{center}
\end{figure}\\dissipation section in the $R(H)$ curves evidence that
the grains have not yet been penetrated at this field range, and
there is no intergranular pinning. Irreversible $R(H)$ loops
appear when the local field at the junctions penetrates the
grains. For polycrystalline high-T$_{c}$ cuprates this local field is higher than $%
H_{ext}$ due to flux compression associated to diamagnetic shielding,
leading to an underestimation of $H_{c1}$. For Ru-1222, the contribution of
the FM magnetization to the local field at the intergranular network, $%
H_{Ru} $, is oriented in opposition to $H_{ext}$, diminishing the flux
compression effect. It was recently determined that $H_{Ru}$ is
approximately 15 Oe at $T $= 0 K for Ru-1222.\cite{das Virgens} Since $%
H_{irr}$ varies from 100 to 3000 Oe, any possible difference between the
local penetration field of the grains and $H_{ext}$ due to the contribution
of $H_{Ru}$ is approximately 10\% or less. Therefore, by determining $%
H_{irr} $ one obtain a good estimation of $H_{c1}$.

Let us consider the effects of the flux trapped in the grains on
the $R(H)$ loops. For a given $H_{max}$ $>$ $H_{c1}$, a fraction
of the grain volume is penetrated. When $H_{ext}$ is decreased
different field profiles are created inside the grains due to
pinning, gradually evolving as $H_{ext}$ diminishes.\cite{Bean}
Each field profile represents a spatial circulation of currents,
i.e., a certain SC magnetization, contributing with
a local field $H_{SC}$ at the intergranular junctions. The lowest value of $%
H_{ext}$ at which full reversal is obtained is $H_{ext}=H_{c1}+2H^{\ast }$.%
\cite{Altshuler} Until the fully reversed profile is not attained, the exact
value of $H_{SC}$ for a given $H_{ext}$ in the returning curve changes as $%
H_{max}$ increases, because a different field profile is obtained for that
external field.

The extent of the compensation between $H_{ext}$ and $H_{SC}$ at the
junctions as the returning curves are collected is clearly evidenced by the
changes in the maxima of $A_{P}(H_{max}),$ the amplitude of the peak in $%
\Delta R(H)$, as shown in Fig. 3. The peaks correspond to a
maximum average compensation of the local fields in the
polycrystal, a relative maximum in the connectivity of the
weak-linked network, as $H_{ext}$ is decreased. The increase in
amplitude of the peaks for higher $H_{max}$\ is associated to a
narrowing of the distribution of local fields in the sample as the
SC magnetization of the grains rises. The position of the peaks is
shifted to higher fields with the rise in $H_{max}$ up to
$H_{sat}$, because the field profile inside the grains becomes
reversed in a larger extent, gradually leading to a higher
$H_{SC}$ in opposition to $H_{ext}$; the maximum compensation
occurs at higher $H_{ext}$. Once the external field is high enough
to reach the fully reversed profile there will be no further
increase in $H_{SC}$ with the rise in $H_{max}$, and the
$A_{P}(H_{max})$ maxima remain unchanged. Therefore
$H_{sat}=H_{c1}+2H^{\ast }$. Since the FM contribution to the
local field is not hysteretic, it has no effect on $H_{sat}$.
\begin{figure}[h!]
 \begin{center}
  \includegraphics[scale=0.54]{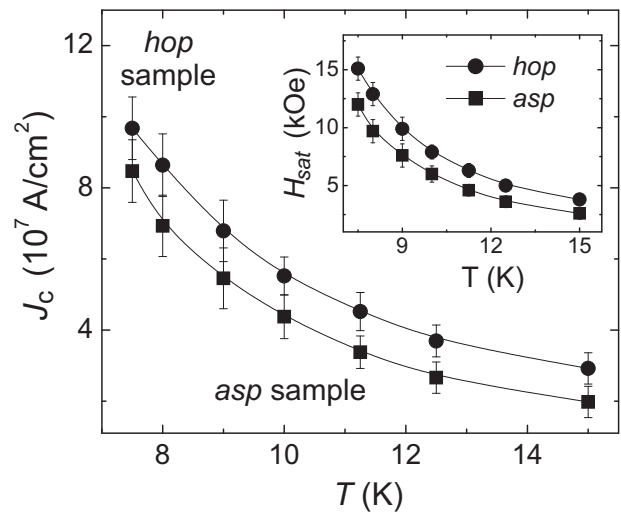}
  \caption{{\small{\it Temperature dependence of the intragrain critical current density $%
J_{c}$ for both Ru-1222 \textit{asp} and \textit{hop} samples. The
inset shows the temperature dependence of the saturation field
$H_{sat}$. The continuous lines are guides to the eyes.}}}
 \end{center}
\end{figure}

Let us compare the absolute values and temperature dependence of $H_{c1}$ in
Ru-1222 with YBCO. Due to the wide SC transition width in the former, the
higher temperature for which $H_{c1}$ was determined is $T$ = 15 K ($%
T/T_{SC} $ $\cong $ 0.3). For this $T/T_{SC}$ value, the lower critical
field in YBCO has already reached its saturation value, $H_{c1}$(YBCO) $%
\cong $ 850 Oe\cite{Wu} (with $H$ parallel to the \textit{c}-axis), while it
is only 150 Oe for both Ru-1222 samples. However, due to its monotonic
increase on cooling, $H_{c1}$ reaches a value twice larger than YBCO for the
Ru-1222 \textit{asp} sample, and even higher for the \textit{hop} one (3000
Oe) at 7.5 K ($T/T_{SC}$ $\cong $ 0.2).

At this point it is important to consider whether there is a possible
contribution from surface barriers in the determination of $H_{c1}$. It
should be noted that Ru-1222 exhibits a monotonic increase of $1/\lambda
^{2}(T)$ on cooling without saturation down to $T$ = 5 K,\cite{Xue02} at
variance with the high-T$_{c}$ cuprates. This implies a reduced efficiency
of any possible barrier shielding with the decrease in temperature, since a
decrease in $\lambda $ favours the flux penetration through weak spots at
the grain surface. Another important point is the strong decrease in $%
\lambda (T)\ $in samples with higher partial oxygen pressure during the
final heat treatment.\cite{Xue02} Therefore, flux penetration through the
weak spots must be easier for the \textit{hop} sample in comparison to the
\textit{asp} one, and the higher $H_{c1}$ values obtained for the former can
not be explained as due to surface barriers. In addition, Ru-1212 and
Ru-1222 present a very distinctive structural feature: the existence of
sharp antiphase boundaries with local distortions separating structural
domains of coherently rotated RuO$_{6}$ octahedra. A high density of quite
sharp boundaries, of several tens of unit cells in length, have been
observed in high resolution transmission electron microscopy images.\cite%
{McLaughlin} In ceramic superconductors, long-range distortions, as twinning
planes\cite{Konczykowski} and columnar defects,\cite{Koshelev} are known to
act as channels for flux penetration, dramatically suppressing surface
barriers shielding. Antiphase boundaries should work very similarly. These
considerations strongly indicate that surface barriers effects are not the
source of the observed $H_{c1}$ dependence in polycrystalline Ru-1222.

The large $H_{c1}$ values obtained at low temperatures are somewhat
unexpected since, as mentioned above, previous reports based on $M(H)$
curves inferred $H_{c1}$ values as low as 25-50 Oe in Ru-1222.\cite%
{Awana,Felner01,Felner02} Nevertheless, clear evidence that these reports
greatly underestimate the magnitude of $H_{c1}$ comes from measurements in
samples with partial substitution of Ru by non-magnetic elements. The
negative diamagnetic minimum in $M(H)\ $is shifted to $\sim $ 300 Oe for
30\% substitution of Ru by Sn,\cite{Williams} and 40\% substitution by Nb,%
\cite{Felner04} even though $T_{SC}$ is decreased in both cases. Moreover,
the zero net magnetization point in the virgin $M(H)$ curves (for which a
significant diamagnetic signal is still present) is observed for $H_{ext}$ $%
\sim $ 800 Oe in both systems. These results indicate that the masking
effect of the FM background is strong, and that the actual magnitude of $%
H_{c1}$ is in fact considerably higher. Further evidence of $H_{c1}\ $values
of the order of several hundreds Oe in ruthenate-cuprates comes from the
magnitude of the internal Ru magnetization and the local fields at different
points of the cell. It has been determined that the in-plane magnetization
at the RuO$_{2}$ layers is $4\pi M\sim 4$ kG, yielding a macroscopic (cell
volume average) dipolar internal field $B_{int}=$\ $4\pi $$<$$M$%
$>$$=700$ Oe.\cite{Pickett} The latter value is confirmed by Gd$%
^{3+}$-electron spin resonance\cite{Fainstein} and muon spin rotation\cite%
{Bernhard01} measurements, giving internal fields of $\sim $ 600-700 Oe at
the Gd site and near to the apical oxygen of the CuO$_{5}$ pyramid,
respectively. Flux expulsion from the grains can occur only if, at a certain
sufficiently low temperature, $H_{c1}(T)$ exceeds the magnetization value of
$4\pi M$. This \textquotedblleft internal\textquotedblright\ Meissner effect
has been previously observed in Ru-1222 at $T$ = 16 K, which is about 30 K
below $T_{SC}$, in samples made using an oxygen flow.\cite{Bernhard02}

The temperature dependence of $H_{c1}$ can be understood when the magnetism
of the Ru sublattice is properly considered. A model for the magnetic
properties of the Ru-1212 system that incorporates the theory of the
superconducting/ferromagnetic multilayers predicts an increase in $H_{c1}$
with the decrease in temperature.\cite{Houzet} This response was obtained
assuming $T_{M}$ $\gg $ $T_{SC}$, which is the case for both Ru-1222 and
Ru-1212 compounds, and ascribed to the decrease of the effective Josephson
coupling between the SC planes due to the exchange field at the FM layers.
In addition, the mentioned $1/\lambda ^{2}(T)$ dependence for Ru-1222
supports the increase of $H_{c1}$ as the temperature diminishes, within the
framework of a BCS-type behavior. Under this assumption, the strong increase
in $1/\lambda ^{2}(T)$ with the rise of the oxygen pressure supports a
significant increase of $H_{c1}$ for the \textit{hop} sample.

The order of magnitude of $J_{c}$ was determined using the expression $%
J_{c}=H^{\ast }/(d/2)$, corresponding to the Bean critical state model for
an infinite SC slab,\cite{Bean} with $d$ $\cong 1$ $\mu m$\ and $H^{\ast
}=(H_{sat}-H_{c1})/2$. The expression for an infinite slab gives the highest
possible values for $J_{c}$ according to the geometry of the SC grain. For a
cylindrical geometry there will be a decrease by one order of magnitude.
Therefore, the calculated values $J_{c}$ $\sim $ 10$^{7}$ A/cm$^{2}$ should
be considered as an upper limit. This large order of magnitude can be
explained in terms of the magnetic phase separation scenario.\cite{Xue04}
Ferromagnetic nanoclusters in the normal state are embedded in an
antiferromagnetic matrix in which superconductivity nucleates. Since the SC
coherence length $\xi $ for Ru-1222 is about the same size as the magnetic
clusters,\cite{Escote} they can effectively act as pinning centers. The
monotonic increase of $J_{c}$ with the decrease in temperature can also be
qualitatively explained in terms of the segregation of FM clusters. It has
been demonstrated for Ru-1222(Eu) that the saturation magnetization of this
magnetic species monotonically increases on cooling down to 10 K,\cite{Xue04}
well inside the SC region. This result\textit{\ }is due to an increase of
the density of FM clusters and/or to variation of their sizes with the
decrease in temperature. As the number of magnetic clusters increases at
lower temperatures, the average matching between the Abrikosov lattice and
the spatial arrangement of pinning-cluster centers is gradually improved.

Finally, it is worth noticing that while oxygenation under pressure
increases the values of $H_{c1}$ by nearly a factor of two at low
temperatures, the increase in $J_{c}$ is moderate. These results suggest
that the oxygen treatment has a larger effect in the region near the grain
boundaries. Flux penetration begins at lower fields for the \textit{asp}
sample because a larger deviation from the oxygen stoichiometry at the grain
borders locally depresses superconductivity. Since $J_{c}$ is influenced by
the properties of the grain as a whole it is less affected by the
oxygenation process. The significant decrease of the absolute resistivity
values and the reduction in the width of the SC transition for the \textit{%
hop} sample in comparison to the \textit{asp} one, while the intragrain SC
transition temperature increases only by $\sim $ 3 K, confirm that the
intergranular connectivity is greatly improved with oxygenation under
pressure, while the core of the grains change in a much less extent.

\section{Conclusions}

In summary, fundamental characteristic properties of Ru-1222, the lower
critical field of the grains and the intragrain critical current density,
were determined after a detailed processing of magnetoresistance loops, $%
R(H) $. The $H_{c1}$ values at low temperatures in well oxygenated
samples are about fifty times larger than the naive estimation
obtained from magnetization measurements at low magnetic fields;
the magnitude of $J_{c}$ is similar to the high-T$_{c}$ cuprates.
The possible segregation of normal-state ferromagnetic
nanoclusters provides a natural scenario to explain the monotonic
increase of both $H_{c1}(T)$ and $J_{c}(T)$\ on
cooling.
\begin{center}\textbf{Acknowledgements}
\end{center}
\hspace{0.2cm}We thank S. Smansunaga and R. Jardim for assistance
with sample oxygenation under high pressures, and E. Altshuler for
useful comments. This work was partially financed by CNPq and
FAPERJ. S.G. was supported by CAPES.

\end{document}